\documentclass[11pt]{amsart}

\usepackage{graphicx}
\usepackage[left=3cm,right=3cm,top=3.25cm,bottom=3.25cm]{geometry}

\usepackage[sort&compress,round]{natbib}
\usepackage{url}

\usepackage{pstricks,pst-node,pst-tree}

\input xy
\xyoption{all}
\CompileMatrices

\newcommand{\tree}[2]{\pstree[treemode=D,arrows=->,treefit=tight,treesep=0.5cm,levelsep=1cm]{#1}{#2}}
\newcommand{\node}[1]{\Toval{#1}}
\renewcommand{\root}[1]{\Toval{#1}}

\newcommand{\cI}{\mathcal{I}}  
\newcommand{\cC}{\mathcal{C}}  
\DeclareMathOperator{\Prob}{Prob}
\newcommand{\pa}{\mathrm{pa}}

\begin{document}

\title[Mutagenetic tree HMM]{A Mutagenetic Tree Hidden Markov Model for
  Longitudinal Clonal HIV Sequence Data}

\author[N. Beerenwinkel]{Niko Beerenwinkel}
\address{Dept.~of Mathematics, University of California,
  1073 Evans Hall, Berkeley, CA 94720}
\email{niko@math.berkeley.edu}

\author[M. Drton]{Mathias Drton}
\address{Dept.~of Statistics, The University of Chicago, 
  5734 S.~University Ave., Chicago, IL 60637}
\email{drton@galton.uchicago.edu}

\begin{abstract}
  RNA viruses provide prominent examples of measurably evolving
  populations. In HIV infection, the development of drug resistance
  is of particular interest, because precise predictions of the outcome of 
  this evolutionary process are a prerequisite for the rational design of antiretroviral
  treatment protocols. We present a
  mutagenetic tree hidden Markov model for the analysis of longitudinal
  clonal sequence data. Using HIV mutation data from clinical trials,
  we estimate the order and rate of occurrence of seven
  amino acid changes that are associated with resistance to the
  reverse transcriptase inhibitor efavirenz.
\end{abstract}

\maketitle

\section{Introduction}

In the developed world, patients infected with human immunodeficiency virus
(HIV) are treated with combinations of several antiretroviral drugs in
order to maximally suppress virus replication. However, the development of
drug resistant virus variants limits the success of this
intervention \citep{Clavel2004}.  The genetic basis for the emergence of
drug resistance is the high mutation rate in HIV, which is a result of the
absence of a proof-reading mechanism for reverse transcription of RNA into
DNA.  The frequent mutations may alter the genetic composition of drug
targets and create drug resistant mutants that will outcompete the wild
type virus and eventually dominate the intra-host virus population.  The
applied drug cocktail then becomes ineffective and the patient experiences
an increase in virus load.  

The accumulation of resistance-conferring mutations is a stochastic process
that typically follows one of several co-existing evolutionary pathways
\citep{Boucher1992,Molla1996,Shafer2005}.  Understanding the
pathways to drug resistance is important for the design of effective drug
combinations.  For the study of these pathways, two types of DNA sequence
data, based on either population or clonal sequencing, are available.

In routine clinical diagnostics, viral drug targets are sequenced after
therapy failure and prior to a therapy switch. Typically, population
sequencing is applied, which results in one DNA sequence that represents
the mixture of viruses in the population. This approach detects only those
mutations that are present in at least 20\% of the strains.  Furthermore,
the phase information, i.e., the knowledge whether polymorphisms at
different sites actually co-occur on the same genome, is lost in population
sequencing. Abundant in public databases \citep{Rhee2003}, cross-sectional
data obtained from population sequencing have been used to demonstrate
correlations between the occurrences of different mutations
\citep{Gonzales2003a,Sing2005a}.  Mutagenetic tree models, a subclass of
Bayesian networks based on directed trees \citep{Beerenwinkel2005f}, take
such correlation analysis a step further by providing a tool particularly
suited to infer evolutionary pathways to drug resistance from
cross-sectional data \citep{Beerenwinkel2005a}.

A more elaborate alternative to population sequencing is clonal sequencing.
In this approach, multiple viral genomes are independently amplified by PCR
and sequenced. This strategy has two major advantages over population
sequencing. First, since mutations may occur in PCR reactions, the use of
multiple independent PCR reactions reduces the impact of early and
erroneous introduction of mutations at this step of the analysis.  Second,
in clonal sequencing, the genotype of a single strain is
determined. From the resulting haplotypes the linkage
between mutations is readily assessed.

While haplotypes provide exact information on the correlation between
mutations, we can gain additional insight about the order in which
substitutions are fixed into the population from longitudinal data obtained
by sequencing viruses from the same patient at multiple time points.  In
this paper, we propose an extension of the mutagenetic tree model to a
model for longitudinal clonal HIV sequence data.  The new model captures
both the time structure per patient and the clonal variation per time
point, and thus allows to employ time structure in the estimation of
substitution rates of resistance-conferring mutations.  Knowledge of these
substitution rates, which can be interpreted as waiting times, provides the
basis for rational therapy planning. Indeed, using viral evolutionary
information encoded in mutagenetic trees has recently been shown to
significantly improve predictions of {\em in vivo} virological response to
antiretroviral combination therapies \citep{Beerenwinkel2005h}.

Rather than modeling viral evolution in full generality, we intend to
describe a specific phase of evolution in terms of a specific set of
genetic events.  More precisely, we model amino acid changes that confer
drug resistance under the constant selective pressure of a given drug.  The
data analyzed in this paper comprise a total of 3350 clones that have been
collected from 163 patients at different time points during three phase II
clinical studies of the reverse transcriptase (RT) inhibitor
efavirenz \citep{Bacheler2000,Bacheler2001}.  We focus on seven amino acid
substitutions in the HIV-1 RT that are associated with resistance to
efavirenz (cf.\ Table~\ref{tab:P22}, where each observed clone is
coded as a binary vector of length seven).  Working on
protease sequences obtained from the same efavirenz studies,
\citet{Foulkes2003b} have modeled evolutionary pathways to drug
resistance by grouping the observed clones into five clusters and
estimating transition rates between clusters.  In that approach, however,
neither the dependence structure between single mutations nor their rates
of occurence is modelled explicitly.  In addition, clonal variation is not
incorporated explicitely.

Under treatment with efavirenz, virus strains harboring one or more of the seven
considered resistance mutations have a significant fitness advantage.  As a
result, these mutations accumulate in the population during the course of therapy
\citep{Boucher1992,Molla1996,Crandall1999a}.  When modelling this
accumulative evolutionary process we make the following three assumptions:
\begin{enumerate}
\item[(A1)] Substitutions do not occur independently. There are preferred
  evolutionary pathways in which mutations are fixed.

\item[(A2)] The fixation of mutations into the populations is definite,
  i.e., substitutions are non-reversible.

\item[(A3)] At each time point, the virus population is dominated by a
  single strain and clones are independent and (possibly erroneous) copies
  of this genotype.
\end{enumerate} 

As mentioned above, previous studies have provided overwhelming evidence
for strong dependencies between resistance mutations, as stated in
assumption (A1).  In fact, the fixation of mutations typically exhibits
order constraints that reflect properties of the underlying fitness
landscape \citep{Beerenwinkel2006a}.  For example, if two
advantageous mutations show positive epistasis, then they interact 
synergistically and the double mutation will be fixed more often than 
expected from the frequencies of the individual substitutions
\citep{Michalakis2004}. In other words, the
mutations are more likely to occur on the same mutational pathway.  On the
other hand, if advantageous mutations exhibit negative epistasis, then they interact 
antagonistically, resulting in mutants that are less fit than expected
or even not viable at all. The mutagenetic tree model
accounts for lethal mutational patterns by excluding some binary vectors
from its state space, which thus encodes only viable mutant types.
Mutational patterns in the state space arise from mutation accumulation
along one of the pathways specified in the tree and we call these patterns
{\em compatible} with the tree.
  
Assumption (A2) refers to the non-reversibility of substitutions. It is
motivated by the strong selective advantage that each resistance mutation
may confer.  Due to this strong selective advantage, we interpret the
fixation of each resistance mutation as a selective sweep, after which
virtually all viable viruses exhibit the mutation.  Together with
assumption (A1), this implies that resistance mutations accumulate along
different pathways by successive sweeps and that both back substitutions
and incompatible states result in viruses whose relative fitness is so low
that they can be ignored.
  
The third assumption (A3) ties in with our point of view of a series of
selective sweeps. Indeed, immediately after such an incident the population
will consist of the new advantageous mutant type and its recent
descendants, which are therefore phylogenetically independent.  Under this
assumption a virus population should exhibit very low genetic diversity.  In
light of the phylogenetic independence, 
the aim of our work is different from that of work employing
population genetic or phylogenetic methods \citep{Drummond2002,Drummond2003}.
Those approaches explicitely model the mutation and selection process  
and typically aim at inferring global population parameters, 
such as the global rate of adaption
\citep{Williamson2003} or the effective population size \citep{Seo2002}.
By contrast, we model the joint probability distribution resulting 
from this evolutionary process directly, with the goal of inferring
individual substitution rates, as the presence of specific mutational
patterns determines the therapeutic options of each patient.

The three assumptions discussed above lead us to a hidden Markov model
(HMM) whose state space consists of the compatible states of a given
mutagenetic tree. We consider unobservable population states and assume
a simple error process involving only a false positive and a false negative
rate that generates clones from this hidden state.  The fixation of each
mutation is modeled by an independent Poisson process. Thus, at each time
point, we assume the virus population to be dominated by a single
mutational pattern and all clonal variation of this unobserved state is
modelled as resulting from erronous copying of this strain.

The remaining part of this paper is organized as follows.  In
Section~\ref{sec:dataassump} we present the data on the seven considered
mutations associated with resistance to efavirenz and perform simple
permutation tests to assess the validity of assumptions (A2) and
(A3).  In Section~\ref{sec:model}, we develop the above outlined
mutagenetic tree HMM (mtree-HMM).  We present the results from fitting the
mtree-HMM to the efavirenz data in Section~\ref{sec:results} and conclude
in Section~\ref{sec:limit} where we discuss some of the limitations of our
approach.

\section{Longitudinal clonal HIV sequence data and model assumption}
\label{sec:dataassump}

\subsection{HIV data}
\label{sec:hiv-data}

We subsequently analyze a set of longitudinal clonal HIV sequences that
have been collected during three clinical studies of efavirenz combination
therapy (DMP 266-003, -004, -005). This data set is publicly available at
the Stanford HIV drug resistance database \citep{Rhee2003}. Selection of
samples, RNA amplification, cloning, and sequencing have been described in
detail by \citet{Bacheler2000} and \citet{Bacheler2001}.  All
patients received the non-nucleoside RT inhibitor efavirenz.

\citet{Bacheler2000} identified a list of amino acid changes in the
HIV RT associated with resistance to efavirenz.  In particular, they
described two alternative pathways to efavirenz resistance, one initiated
by mutation K103N (103-pathway), the other by Y188L (188-pathway).  In the
present data set, for each patient, at most one of these two pathways
occurs. We focus here on the 103-pathway, because mutations associated with
the 188-pathway were found in only 7 out of 170 patients, which we excluded
from further consideration.  A total of $3350$ clones have been derived
from the remaining $163$ patients at between 1 and 11 different time points
(median = 3, IQR 1--5).  From the list of efavirenz resistance mutations
associated with the 103-pathway we selected those that occur in at least
$2\%$ of clones. This strategy resulted in the seven mutations 103N
(frequency 48.1\%), 225H (8.0\%), 108I (4.9\%), 101Q (2.7\%), 190S (2.7\%),
101E (2.5\%), and 100I (2.3\%).

\begin{table}[!t]
\small
\begin{tabular}{ccccccccc} \hline
Week & 100I  & 101E  & 101Q  & 103N  & 108I  & 190S  & 225H  \\ \hline
   0 & \em 0 & \em 0 & \em 0 & \em 0 & \em 0 & \em 0 & \em 0 \\
& 0  & 0  & 0  & 0  & 0  & 0  & 0 \\
& 0  & 0  & 0  & 0  & 0  & 0  & 0\\
& 0  & 0  & 0  & 0  & 0  & 0  & 0\\
& 0  & 0  & 0  & 0  & 0  & 0  & 0\\
& 0  & 0  & 0  & 0  & 0  & 0  & 0\\
& 0  & 0  & 0  & 0  & 0  & 0  & 0\\
& 0  & 0  & 0  & 0  & 0  & 0  & 0\\
& 0  & 0  & 0  & 1  & 0  & 0  & 0\\
& 0  & 0  & 0  & 0  & 0  & 0  & 0\\
& 0  & 0  & 0  & 0  & 0  & 0  & 0\\
& 0  & 0  & 0  & 0  & 0  & 0  & 0\\
& 0  & 0  & 0  & 1  & 0  & 0  & 0\\
& 0  & 0  & 0  & 0  & 0  & 0  & 0\\
& 0  & 0  & 0  & 0  & 0  & 0  & 0\\
& 0  & 0  & 0  & 0  & 0  & 0  & 0\\
& 0  & 0  & 0  & 0  & 0  & 0  & 0\\\hline
48 & \em 0 & \em 0 & \em 0 & \em 1 & \em 0 & \em 0 & \em 0 \\
& 0 &  0 &  0 & 1 &  0 &  0 & 0\\
& 0 &  0 &  0 & 1 &  0 &  0 & 0\\
& 0 &  0 &  0 & 1 &  0 &  0 & 0\\
& 0 &  0 &  0 & 1 &  0 &  0 & 0\\\hline
59 & \em 0 & \em 0 & \em 0 & \em 1 & \em 0 & \em 0 & \em 0\\
& 0 &  0 &  0 & 1 &  0 &  0 & 0\\
& 0 &  0 &  0 & 1 &  0 &  0 & 0\\
& 0 &  0 &  0 & 1 &  0 &  0 & 0\\
& 0 &  0 &  0 & 1 &  0 &  0 & 0\\
& 0 &  0 &  0 & 1 &  0 &  0 & 0\\
& 0 &  0 &  0 & 1 &  0 &  0 & 1\\
& 0 &  0 &  0 & 1 &  0 &  0 & 0\\\hline
70 & \em 0 & \em 0 & \em 0 & \em 1 & \em 0 & \em 0 & \em 1\\
& 0 &  0 &  0 & 1 &  0 &  0 & 1\\
& 0 &  0 &  0 & 1 &  0 &  0 & 1\\
& 0 &  0 &  0 & 1 &  0 &  0 & 0\\
& 0 &  0 &  0 & 1 &  0 &  0 & 0\\
& 0 &  0 &  0 & 1 &  0 &  0 & 0\\
& 0 &  0 &  0 & 1 &  0 &  0 & 0\\
& 0 &  0 &  0 & 1 &  0 &  0 & 1\\
& 0 &  0 &  0 & 1 &  0 &  0 & 1\\\hline
~\\
\end{tabular}
\caption{Mutation data and inferred Viterbi path for patient 22. Presence
   and absence of mutations are encoded as ``1'' and ``0'',
   respectively. At each time point (week) the first row represents the
   inferred population state (in italics), all other rows are unordered
   and correspond to observed clones.} 
\label{tab:P22}
\end{table}

As an example for mutational patterns found in the data set, 
Table~\ref{tab:P22} shows the clones observed in patient 22, which is also highlighted
in \citet[Tab.~2]{Bacheler2000} (Viterbi paths such as the one given
in Table~\ref{tab:P22} are discussed in Section~\ref{sec:results}).  This
patient is atypical in that before the onset of therapy (week 0), two
clones had mutation 103N.  In fact, among the 1144 baseline clones only
very few already featured the mutations considered here, i.e., mutations
103N (frequency 1.9\%), 190S (1.3\%), 101Q (0.7\%), 101E (0.35\%), and 108I
(0.08\%).  Mutations 100I and 225H were not present in any baseline clone.
Only six clones (0.52\%) harbored a double mutation, namely 103N+190S.

According to Table~\ref{tab:P22}, the first selective sweep, introducing
mutation 103N in the virus population of patient 22, has occurred by week
48.  By week 70, mutation 225H shows definite signs of manifestation.  This
behaviour is in support of the assumptions (A2) and (A3) put forward in the
Introduction.  For a quantitative analysis of these assumptions, we perform
randomization tests.

\subsection{Statistical tests}
\label{sec:statistical-tests}

Let $N=163$ be the number of patients in our data set, and let $K_{ij}$ be
the number of clones observed in patient $i\in [N]:=\{1,\dots,N\}$ at the
$j$-th time point $t_{ij}$.  Moreover, let $y_{ijkm}\in\{0,1\}$ be the
indicator of the presence of mutation $m\in [M]:=\{1,\dots,7\}$ in clone
$k\in [K_{ij}]$ derived from patient $i\in [N]$ at time point $t_{ij}$.  We
use two test statistics for evaluating the assumptions (A2) and (A3),
respectively. In both cases, we treat different patients as independent
observations. 

The non-reversibility of substitutions (A2) is tested by tracing the change
in mutant allele frequency for each patient over time.  This frequency is
\[
   f_{i,j,m} = \frac{1}{K_{ij}} \sum_{k \in K_{ij}} y_{ijkm},
     \quad i \in [N],\: j \in J_i,\: m \in [M].
\]
For each mutation $m$, our test statistic $A_m$ counts how often 
its frequency decreases from one time point to the next. 
Thus,
\[
   A_m = \frac{1}{N} \sum_{i \in [N]} \frac{1}{J_i - 1} 
           \sum_{j=1}^{J_i - 1} I \{ f_{i,j,m} > f_{i,j+1,m} \},
\]
where $I$ denotes the indicator function. We estimate the distribution of
$A_m$ under the null hypothesis of equal probability of increase and
decrease of allele frequencies by randomizing the temporal order of the
clone populations.  Under non-reversibility of substitutions, the observed
value of the test statistic $A_m$ should be small compared to the
randomization-based values of $A_m$.


\begin{figure}[!tpb]
\centering
\includegraphics[width=7cm]{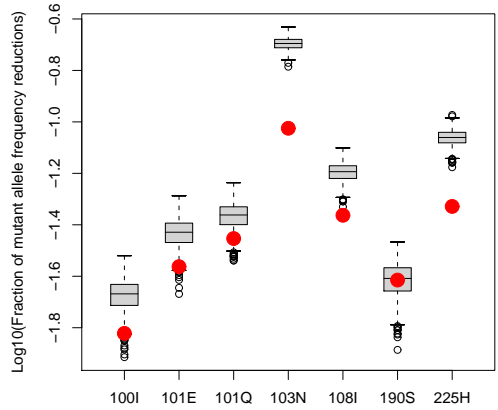}
\caption{Non-reversibility of substitutions. For each mutation,
the logarithm of the number of times for which its frequency in the 
population strictly decreases between successive time points 
is displayed (bold discs). The distribution of this test statistic was
estimated by shuffling 1000 times the order of time points 
(box plots).
}
\label{fig:accumulative}
\end{figure}

In our data, mutation losses occur in only 1.5 to 9.4\% of the cases
(Figure~\ref{fig:accumulative}).  In other words, for all seven considered
mutations the frequency of the mutant increases or stays constant in over
90\% of pairs of subsequent virus populations.  For most mutations, this
value was significantly smaller than expected ($p_{\rm 100I} = 0.011$,
$p_{\rm 101E} = 0.013$, $p_{\rm 103N} < 0.001$, $p_{\rm 108I} < 0.001$,
$p_{\rm 225H} < 0.001$) with the exceptions of mutations 101Q ($p_{\rm
  101Q} = 0.056$) and 190S ($p_{\rm 190S} = 0.465$).  We conclude that
assumption (A2) of non-reversible substitutions is reasonable for the
majority of the data.

For assessing the validity of assumption (A3), we measure the genetic
diversity among mutational patterns by the Hamming distance. If $c$ and
$c'$ are two clones, their Hamming distance is defined as
\[
   D_H(c,c') = \sum_{m \in [M]} I \{ c_m \not= c'_m \}.
\]
The diversity of a population of clones $c_1, \dots, c_K$ 
is the expected Hamming distance between two clones drawn 
at random from that population,
\[
   D_H(c_1, \dots, c_K) = \frac{K(K-1)}{2} \sum_{k < k'} D_H(c_k, c_{k'}).
\]
Our test statistic is the expected population diversity
\[
   B = \frac{1}{N} \sum_{i \in [N]} \frac{1}{J_i} \sum_{j=1}^{J_i}
       D_H(y_{ij1}, \dots, y_{ijK_{ij}}).
\]
We consider the null hypothesis of observing the full diversity present in
all clones from one patient at each single time point. The distribution of
$B$ under this null is estimated by randomly permuting the assignment of
clones to time points.  Observed values at the left tail of this
distribution indicate less genetic diversity per time point than expected
from the total diversity present in the clones.  Large population
diversities would put into question our assumption of a single population
state.

\begin{figure}[!tpb]
\centering
\includegraphics[width=7cm]{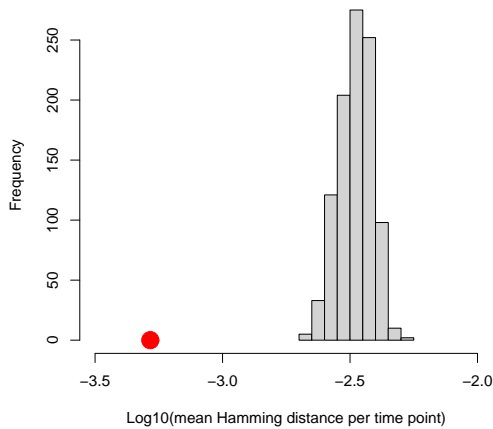}
\caption{Genetic diversity. The logarithm of the average Hamming 
distance between clones from the same time point is displayed 
(bold disc). The distribution of this test statistic was estimated 
by randomizing 1000 times the assignment of clones to time points 
(histogram).}
\label{fig:homogeneous}
\end{figure}

For our data, grouping clones according to their sampling time reduces
the genetic diversity more than expected from a random grouping.  We find
that on average clones from the same population differ in only 1 out of
1900 of the considered mutations (Figure~\ref{fig:homogeneous}). For random
clone groupings the expected value is 1 in 300. This difference is highly
significant ($p < 0.001$).  Thus, most genetic diversity occurs
between time points rather than within time points, and most of the
observed populations indeed look like having gone through a selective
sweep. We therefore conclude that our assumptions described in the
Introduction and formalized in the mutagenetic tree HMM, which is described next, are
in reasonable agreement with the clonal HIV mutation  data
considered here.

\section{Mutagenetic tree hidden Markov model}
\label{sec:model}

We begin this section by recalling the basic definition and properties of
the mutagenetic tree model for cross-sectional data. We first extend this
basic model to a mutagenetic tree Markov chain model for multiple
observations per patient in time. Second, we introduce the mutagenetic tree
hidden Markov model (mtree-HMM), which assumes the evolutionary state of
the virus population to be unobservable.  An EM algorithm for parameter
estimation is presented, the details of which are given in the Appendix.

\subsection{Mutagenetic tree model}

Consider $M$ genetic events of interest, which we also call
\emph{mutations}.  We treat these events as random and use binary random
variables to indicate their occurrence.  For each mutation $m\in [M] :=
\{1,\dots,M\}$, let $X_m$ be a binary random variable taking values in
$\{0,1\}$ such that $\{X_m=1\}$ and $\{X_m=0\}$ indicate the occurrence and
non-occurrence of mutation $m$, respectively.  In the HIV application,
$[M]$ represents the set of seven amino acid changes in the HIV RT
described in Section~2 that
confer resistance to efavirenz, and $X_m$ indicates the fixation
of mutation $m$ into the virus population.  For convenience, we
additionally introduce a random variable $X_0$ with $\Prob(X_0=1)=1$.  In
this setup, the mutational pattern representing the evolutionary state of
the virus population corresponds to the binary vector $X = (X_1, \dots,
X_M)$ that takes values in the state space $\cI = \{0,1\}^M$.

A \emph{mutagenetic tree} $T$ on $M$ events is a connected branching on the
set of vertices $V=\{0\}\cup [M]$, rooted at node $0$
\citep{Beerenwinkel2005f}; in the cancer literature, such a tree
has been termed oncogenetic \citep{Desper1999,vonHeydebreck2004}.  In particular, this
tree is an acyclic directed graph and thus induces a directed graphical
model for the joint distribution of the random vector $X$
\citep{Lauritzen1996}.  The joint distributions for $X$ comprised in this
graphical model factor as
\[
\Prob(X=x)=
\prod_{m\in [M]}
 \Prob(X_m = x_m \mid X_{\pa(m)} = x_{\pa(m)}), 
\]
where $x\in\cI$, $x_0:=1$, and $\pa(m)=\pa_T(m)$ denotes the
parent of $m$ in $T$.
The \emph{mutagenetic
  tree model} induced by the tree $T$ is a submodel of the usual
graphical model obtained by restricting the conditional probability
matrices to the form
\begin{equation*}
\vartheta^m =
\bordermatrix{   & 0                  & 1 \cr
               0 & 1                  & 0 \cr
               1 & 1-\vartheta^m_{11} & \vartheta^m_{11}
} 
\end{equation*}
where $\vartheta^m_{ab} = \Prob(X_m = b \mid X_{\pa(m)} = a) \in [0,1]$.
The mutagenetic tree model imposes the same conditional independence
structure as the graphical model but in addition imposes the constraint
that an event can occur only if all of its ancestor events have already
occurred \citep[Thm.~14.6]{Beerenwinkel2005c}.  This fact follows from the
first row of the transition matrix.  As a consequence, mutations can only
occur in specific pathways.  More precisely, each mutational pattern
corresponds to a subtree of the mutagenetic tree and the pathways
correspond to sequences of subtrees related by inclusion and increasing by
exactly one vertex (mutation).  The extreme cases are the star and the
path.  In the star, each vertex has the root vertex as parent and all
mutational pathways are possible.  In the path, the vertices form a chain
starting with the root vertex.  In this case, there is only one pathway.

\begin{figure}[!tpb]
\centering
\begin{tabular}{lr} 
(a) \tree{\root{$X_0$}}{
        \tree{\node{$X_1$}}{
                \node{$X_2$}
                \tree{\node{$X_3$}}{
                        \node{$X_4$}
                        }
                }
        }
\newlength{\MyLength}
\settowidth{\MyLength}{$X$}
\newcommand{\myNode}[2]{\circlenode{#1}{\makebox[\MyLength]{#2}}}
\psset{linewidth=0.75pt}          
& (b) \xymatrix@!=0cm{
&1111\\
1011\ar@{-}[ur] &&  1110\ar@{-}[ul] &\\
&1010\ar@{-}[ur]\ar@{-}[ul] && 1100\ar@{-}[ul]\\
&& 1000\ar@{-}[ur]\ar@{-}[ul]\\
& 0000\ar@{-}[ur] &&   }
\end{tabular}
\caption{\label{fig:lattice} 
(a) Mutagenetic tree on five vertices; and
(b) its induced lattice of compatible states.
}
\end{figure}
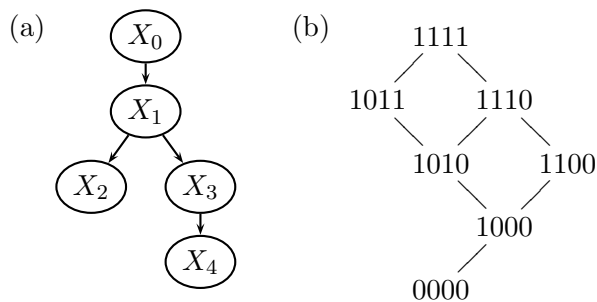

A mutational pattern $x \in \cI$ is called \emph{compatible} with
the tree $T$, if $x$ is a state that may occur with non-zero
probability in the mutagenetic tree model induced by $T$.  Hence, 
$x$ is compatible with $T$, if there exist parameters
$\vartheta = (\vartheta^1_{11}, \dots, \vartheta^M_{11})\in
[0,1]^M$ such that 
\[
\Prob(X=x)=\prod_{m\in [M]}
\vartheta^m_{x_{\pa(m)},x_m}>0.
\]
The set $\cC(T)$ of states that are compatible with the tree $T$ forms a
finite distributive lattice $(\cC(T),\vee,\wedge)$, where $\vee$ and
$\wedge$ denote the component-wise maximum and minimum operator,
respectively \citep[Lem.~14.3]{Beerenwinkel2005c}, cf.\ 
Figure~\ref{fig:lattice}(a),(b).  The so-called Hasse diagram in
Figure~\ref{fig:lattice}(b) illustrates the above-mentioned
pathways in that any such pathway corresponds to a path from the
(wild-type) state $(0,\dots,0)\in\cC(T)$ to the state
$(1,\dots,1)\in\cC(T)$ with all mutations present.

In a \emph{timed mutagenetic tree model}, mutations occur according to
independent Poisson processes.  If $\lambda_m>0$
denotes the rate of this process on the edge $\pa(m)\to m$, then the
probability that mutation $m$ occurs during a time period of length
$\Delta t$ is
\[ 
\vartheta^m_{11} 
= \Prob(X_m = 1 \mid X_{\pa(m)} = 1) 
= 1 - e^{-\lambda_m \, \Delta t}.
\]

Mutagenetic trees can be reconstructed from cross-sectional data by solving
the maximum weight branching problem in the complete graph on the set of
vertices $V$ with an efficient combinatorial algorithm \citep{Desper1999}.
This algorithm is implemented, for example, in the {\tt Mtreemix} software
package \citep{Beerenwinkel2005b}.  Here, we use the algorithm for obtaining
the topology (but not the parameters) of the tree.  For this purpose, the
longitudinal data were treated as cross-sectional.

\subsection{Markov chain model}
\label{sec:markovchain}

Assume now that we can observe the viral mutational pattern within one
patient at more than one time point.  More precisely, let $X_{jm}$ be a
binary random variable indicating the occurrence of mutation $m$ at time
point $t_{j}$, $j=1,\dots,J$, in the patient's virus population.  We assume
that the evolutionary process starts at time 0 in the wild type state,
i.e.\ without any mutation, which is the case for the majority of our data
(cf.\ Section~\ref{sec:hiv-data}).  Thus, the initial population state
distribution follows the timed mutagenetic tree model. In particular,
$X_{1m}=0$ for all $m\in [M]$ with $t_1 = 0$.

For the temporal evolution of the mutational patterns
$X_{j}=(X_{j1},\dots,X_{jM})$, $j=1,\dots,J$, we make the assumption that
these vectors form a Markov chain observed at fixed time points
$t_1<t_2<\dots <t_J$.  Furthermore, we assume that a mutation occurs at the
time it is observed.  In the HIV application, observations can only be made
if the virus load is above a detectable limit resulting in
censored observations.  However, prior work of \citet{Foulkes2003b}
suggests that the impact of the censoring is minor (cf.\
Section~\ref{sec:limit}).

The development of mutation $m$ at a time point $t_{j}$ with $j\ge
2$ which is encoded in the random variable $X_{jm}$, depends on past
mutational events as well as current ancestral mutational events only
through two indicators.  The first one is $X_{j-1,m}$ and indicates whether
the mutation was already present at the immediately preceding time point
$t_{j-1}$. The second one is $X_{j\pa(m)}$ and indicates whether the parent
mutation is present at time point $t_{j}$.  These dependencies arise from
modeling the presence of mutation $m$ at time $t_{j}$ as resulting either
from its introduction along the edge $\pa(m)\to m$ at time point $t_{j}$,
or from its earlier non-reversible introduction into the population and
hence its presence at time point $t_{j-1}$.  The dependency structure among
the variables $(X_{jm}\mid j=1,\dots,J,\; m\in [M])$ can be encoded in an
acyclic directed graph.  For example, the subgraph of the graph shown in
Figure~\ref{fig:mtreehmm} induced by the unshaded vertices represents the
mutagenetic tree Markov chain based on the tree in
Figure~\ref{fig:lattice}(a).

The dynamics of this Markov chain model are encoded by the transition matrices
\begin{equation}
  \label{eq:transtheta}
    \theta_{j}(\lambda_m)     
    = \bordermatrix{ & 0 & 1\cr 
      00 & 1 & 0\cr 
      01 &e^{ -\lambda_m (t_{j}-t_{j-1}) } & 
      1-e^{ -\lambda_m (t_{j}-t_{j-1})}\cr
      10 & * & *\cr 11 & 0 & 1 }
\end{equation}
whose rows are indexed by pairs $(m,\pa(m))$ in $\{0,1\}^2$.  
The asterisks indicate
entries that need not be specified, because 
no event $m$ can occur before its parent event $\pa(m)$.
With these matrices
we define the \emph{mutagenetic tree Markov chain model} as the family of
joint distributions of the form
\begin{equation*}
\Prob(X_{jm}=x_{jm},\; j=1,\dots,J,\; m\in [M])
= \prod_{j=1}^{J} \prod_{m\in [M]} 
\theta_{j}(\lambda_m)_{(x_{j-1,m},x_{j\pa(m)}),x_{jm}},
\end{equation*}
where $x_{j0}:=1$ and $t_0:=0$.
It follows that
\begin{equation*}
\Prob(X_{jm}=x_{jm} \mid X_{j-1,m}=x_{j-1,m},\;
  X_{j\pa(m)}=x_{j\pa(m)}) 
= \theta_j(\lambda_m)_{(x_{j-1,m},x_{j\pa(m)}),x_{jm}}.
\end{equation*}

\subsection{Mtree-HMM}

\begin{figure*}[!tpb]
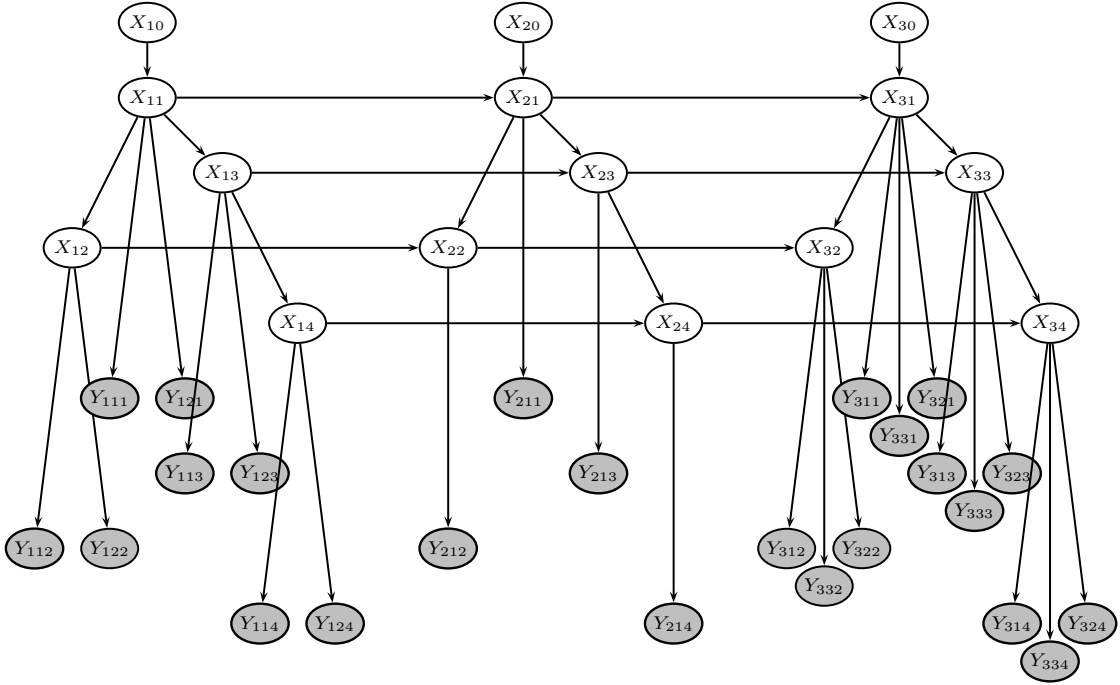

  \centering
    \vspace{10cm}\hspace{-14cm}
    \tiny
    \psset{linewidth=0.75pt}          
    \settowidth{\MyLength}{$X_{1}$}
    \newcommand{\myNode}[2]{\ovalnode{#1}{\makebox[\MyLength]{#2}}}
      \rput(1.5, 9){\myNode{10}{$X_{10}$}} 
      \rput(1.5, 8){\myNode{11}{$X_{11}$}} 
      \rput(0.5, 6){\myNode{12}{$X_{12}$}}
      \rput(2.5, 7){\myNode{13}{$X_{13}$}}
      \rput(3.5, 5){\myNode{14}{$X_{14}$}} 
      \rput(6.5, 9){\myNode{20}{$X_{20}$}} 
      \rput(6.5, 8){\myNode{21}{$X_{21}$}} 
      \rput(5.5, 6){\myNode{22}{$X_{22}$}}
      \rput(7.5, 7){\myNode{23}{$X_{23}$}}
      \rput(8.5, 5){\myNode{24}{$X_{24}$}} 
      \rput(11.5, 9){\myNode{30}{$X_{30}$}} 
      \rput(11.5, 8){\myNode{31}{$X_{31}$}} 
      \rput(10.5, 6){\myNode{32}{$X_{32}$}}
      \rput(12.5, 7){\myNode{33}{$X_{33}$}}
      \rput(13.5, 5){\myNode{34}{$X_{34}$}} 
      \psset{fillstyle=solid}
      \psset{fillcolor=lightgray}
      \rput(1, 2){\myNode{c122}{$Y_{122}$}}
      \rput(10, 2){\myNode{c32}{$Y_{312}$}}
      \rput(11, 2){\myNode{c322}{$Y_{322}$}}
      \rput(10.5, 1.5){\myNode{c332}{$Y_{332}$}}
      \psset{linecolor=black}
      \psset{linewidth=0.75pt}          
      \ncline{->}{12}{c122}
      \ncline{->}{32}{c32}
      \ncline{->}{32}{c322}
      \ncline{->}{32}{c332}
      \psset{linewidth=0.9pt}          
      \rput(1, 4){\myNode{c11}{$Y_{111}$}} 
      \rput(0, 2){\myNode{c12}{$Y_{112}$}}
      \rput(2, 3){\myNode{c13}{$Y_{113}$}}
      \rput(3, 1){\myNode{c14}{$Y_{114}$}} 
      \rput(2, 4){\myNode{c121}{$Y_{121}$}} 
      \rput(3, 3){\myNode{c123}{$Y_{123}$}}
      \rput(4, 1){\myNode{c124}{$Y_{124}$}} 
      \rput(6.5, 4){\myNode{c21}{$Y_{211}$}} 
      \rput(5.5, 2){\myNode{c22}{$Y_{212}$}}
      \rput(7.5, 3){\myNode{c23}{$Y_{213}$}}
      \rput(8.5, 1){\myNode{c24}{$Y_{214}$}} 
      \rput(11, 4){\myNode{c31}{$Y_{311}$}} 
      \rput(12, 3){\myNode{c33}{$Y_{313}$}}
      \rput(13, 1){\myNode{c34}{$Y_{314}$}} 
      \rput(12, 4){\myNode{c321}{$Y_{321}$}} 
      \rput(13, 3){\myNode{c323}{$Y_{323}$}}
      \rput(14, 1){\myNode{c324}{$Y_{324}$}} 
      \rput(11.5, 3.5){\myNode{c331}{$Y_{331}$}} 
      \rput(12.5, 2.5){\myNode{c333}{$Y_{333}$}}
      \rput(13.5, 0.5){\myNode{c334}{$Y_{334}$}} 
      \psset{fillstyle=none}
      \psset{linecolor=black}
      \psset{linewidth=0.75pt}          
      \ncline{->}{10}{11}
      \ncline{->}{20}{21}
      \ncline{->}{30}{31}
      \ncline{->}{21}{22}
      \ncline{->}{31}{32}
      \ncline{->}{12}{22}
      \ncline{->}{22}{32}
      \ncline{->}{11}{c121}
      \ncline{->}{13}{c123}
      \ncline{->}{14}{c124}
      \ncline{->}{21}{c21}
      \ncline{->}{22}{c22}
      \ncline{->}{23}{c23}
      \ncline{->}{31}{c31}
      \ncline{->}{33}{c33}
      \ncline{->}{31}{c321}
      \ncline{->}{33}{c323}
      \ncline{->}{31}{c331}
      \ncline{->}{33}{c333}
      \psset{linecolor=black}
      \psset{linewidth=0.75pt}          
      \ncline{->}{11}{12}
      \ncline{->}{11}{13}
      \ncline{->}{13}{14}
      \ncline{->}{11}{c11}
      \ncline{->}{12}{c12}
      \ncline{->}{13}{c13}
      \ncline{->}{14}{c14}
      \ncline{->}{21}{23}
      \ncline{->}{23}{24}
      \ncline{->}{24}{c24}
      \ncline{->}{11}{21}
      \ncline{->}{13}{23}
      \ncline{->}{14}{24}
      \ncline{->}{31}{33}
      \ncline{->}{33}{34}
      \ncline{->}{21}{31}
      \ncline{->}{23}{33}
      \ncline{->}{24}{34}
      \ncline{->}{34}{c34}
      \ncline{->}{34}{c324}
      \ncline{->}{34}{c334}
    \caption{Acyclic directed graph representing an mtree-HMM
      based on the tree from Figure~\ref{fig:lattice}(a) with
      $J=3$ time points, two clones at time point $t_1$, one
      clone at $t_2$, and three clones at $t_3$. Clear vertices
      correspond to hidden random variables, shaded vertices to
      observable variables.}
    \label{fig:mtreehmm}
\end{figure*}

We now turn to the case of observed clonal samples, rather than population
states, at different time points.  We model these data by assuming that the
clones are erroneous copies of a hidden mutational pattern that evolves
according to a mutagenetic tree Markov chain.  Hence, $X_{jm}$ is now a
hidden binary random variable.  The observed data are instances of binary
random variables $Y_{jkm}$, $k\in [K_{j}]$ that indicate whether mutation
$m$ is present in the $k$-th clone sampled from the virus population at
time point $t_{j}$.  Clones are conditionally independent given the hidden
states $(X_{j})_{j=1,\dots,J}$.  The resulting graphical model is referred
to as the {\em mutagenetic tree hidden Markov model}, or {\em mtree-HMM}
for short.  Its graph is shown in Figure~\ref{fig:mtreehmm}.

Let $\varepsilon^+ = (\varepsilon^+_1, \dots, \varepsilon^+_M) \in [0,1]^M$
and $\varepsilon^- = (\varepsilon^-_1, \dots, \varepsilon^-_M) \in [0,1]^M$
be parameter vectors that contain the mutation-specific probabilities of
observing a false positive and a false negative, respectively. False
positives (negatives) are mutations (wild type residues) observed in clones
derived from a virus population that is in the wild type (mutant) state at
that time point.  The false positive and false negative rates summarize
differences from the population state that can arise from mutations in the
PCR reactions, or from erroneous viral copying of the dominating strain.
Thus, these parameters quantify the expected genetic diversity of the virus
population.  Conditionally upon the hidden state $X_{jm}$, the
probabilities of observing mutation $m$ in clone $k$ at time point $t_{j}$
are
\begin{equation*}  
  \theta'(\varepsilon^+_m,\varepsilon^-_m)  
  = \bordermatrix{
    & 0 & 1\cr
    0& 1-\varepsilon^+_m & \varepsilon^+_m\cr
    1& \varepsilon^-_m & 1-\varepsilon^-_m
  }.    
\end{equation*}
The entries of this matrix are the conditional probabilities
\[
   \theta'(\varepsilon^+_m,\varepsilon^-_m)_{x_{jm},y_{jkm}}  
  = \Prob(Y_{jkm}=y_{jkm} \mid X_{jm}=x_{jm}).
\]
The different clones $Y_{jk}=(Y_{jk1},\dots,Y_{jkm})$, $k\in [K_{j}]$,
are thus modeled as independent and identically distributed.  Let
\[ 
Y = (Y_{jkm} \mid j=1,\dots,J,\;
k\in [K_{j}],\; m\in [M])
\]
denote all clonal sequence observations.  The \emph{mutagenetic tree
  hidden Markov model (mtree-HMM)} is the family of joint
distributions of $Y$ given by
\begin{multline*}  
  \Prob(Y=y) = \\
  \sum_{x_{1} \in \cC(T)} \dots \sum_{x_{J} \in \cC(T)}
  \prod_{m\in [M]}
    \prod_{j=1}^{J}
    \bigg(
      \theta_j(\lambda_m)_{(x_{j-1,m},x_{j\pa(m)}),x_{jm}}
      \prod_{k\in [K_{j}]}
      \theta'(\varepsilon^+_m,\varepsilon^-_m)_{x_{jm},y_{jkm}}
    \bigg),
\end{multline*}
where we have summed over the hidden states.  The graphical model
structure of the mtree-HMM is illustrated in Figure
\ref{fig:mtreehmm}. 

We remark that on an abstract level mtree-HMMs are related to phylogenetic
HMMs (phylo-HMMs) \citep{McAuliffe2004,Siepel2004}.  Phylo-HMMs are
probabilistic models of sequence evolution that combine phylogenetic trees
at different sites of aligned genomes with an HMM that allows for
dependence across sites.  Thus, in phylo-HMMs the tree models arise through
dependence in time, and the HMM represents a dependence in space (along the
genome).  While mtree-HMMs are similarly structured in that tree-based
models are combined with HMMs, the roles of space and time are reversed.
The mutagenetic trees arise from dependence among genome sites, and the HMM
introduces dependence across time.

\subsection{Parameter estimation}   \label{sec:statinf}

Consider now clonal sequence data for a set of patients
indexed by $[N]=\{1,\dots,N\}$.  For each patient $i\in [N]$, we have
observations at time points $t_{i1}<t_{i2}<\dots<t_{iJ_i}$.  We denote
by $X_{ijm}$ the indicator of occurrence of mutation $m$ in the
viral population state of patient $i$ at time point $t_{ij}$.  The
random variable $Y_{ijkm}$ indicates the occurrence of mutation $m$ in
clone $k\in [K_{ij}]$ of patient $i$ at time point $t_{ij}$.
We denote the transition matrices (Equation~\ref{eq:transtheta}) by
$\theta_{ij}(\lambda_m)$. For example,
\[
\theta_{ij}(\lambda_m)_{01,0}=
e^{ -\lambda_m (t_{ij}-t_{i,j-1}) }.
\]

We assume patients to be independent and each patient to follow the
mtree-HMM based on a fixed tree $T$.  Then the resulting model
for the observations
\[ 
Y = (Y_{ijkv} \mid i\in [N],\; j=1,\dots,J_i,\; k\in [K_{ij}],\; m\in
[M] ),
\]
made at time points $\{t_{ij} \mid i\in [N],\; j=1,\dots,J_i \}$, has
the likelihood function
\begin{multline*}
  L_{\rm obs}(\lambda,\varepsilon^+,\varepsilon^-) = 
  \sum_{x_{11} \in \cC(T)} \dots \sum_{x_{NJ_N} \in \cC(T)}
  \prod_{i\in [N]} \prod_{m\in [M]}
  \prod_{j=1}^{J_i} \\
  \bigg(
  \theta_{ij}(\lambda_m)_{(x_{i,j-1,m},x_{ij\pa(m)}),
    x_{ijm}}
  \prod_{k\in [K_{ij}]}
  \theta'(\varepsilon^+_m,\varepsilon^-_m)_{x_{ijm},y_{ijkm}}
  \bigg),
\end{multline*}
where we set $x_{i0m}:=1$
and $t_{i0}:=0$ for all $i \in [N]$ and $m \in [M]$.
The outer sums make the likelihood function hard
to maximize.  In order to solve this optimization problem and to obtain
maximum likelihood estimates, we apply an Expectation Maximization
(EM) algorithm. 

Let $\{x_{ijm}\}$ be values of the hidden variables $\{X_{ijm}\}$ that are
compatible with the unobserved mutagenetic tree Markov chain model.  Then
it must hold that $x_{ijm} = 1$ whenever $x_{i,j-1,m}=1$ and $x_{ijm} = 0$
whenever $x_{ij\pa(m)} = 0$. Let $I$ be the indicator function and set
\begin{eqnarray*}
\chi_{ijm}(a) & = & 
I\{x_{i,j-1,m}=0,\; x_{ij\pa(m)}=1,\; x_{ijm}=a\}, \\
\chi'_{ijkm}(a,b) & = & 
I\{x_{ijm}=a,\; y_{ijkm}=b\},
\end{eqnarray*}
for $a,b=0,1$. The log-likelihood function 
$\ell_{\rm hid}(\lambda,\varepsilon^+,\varepsilon^-)$
of the hidden model is the sum over all
$i \in [N]$, $m \in M$, and $j=1,\dots,J_i$
of the expressions
\begin{multline*}
    - \chi_{ijm}(0) \, \lambda_m \, (t_{ij} - t_{i,j-1})
    + \chi_{ijm}(1) \, \log\big(1 - e^{\lambda_m (t_{ij} -
      t_{i,j-1})}\big) 
    \\
    +
     \sum_{k\in [K_{ij}]}
      \Big[ 
      \chi'_{ijkm}(0,0) \, \log(1 - \varepsilon^+_m)
      + \chi'_{ijkm}(0,1) \, \log\, \varepsilon^+_m \\ +
      \;\,\chi'_{ijkm}(1,0) \, \log\, \varepsilon^-_m
      + \chi'_{ijkm}(1,1) \, \log(1 - \varepsilon^-_m) \Big].
\end{multline*}
Maximization of $\ell_{\rm hid}$ is feasible since it does not involve
the sums that are present in $L_{\rm obs}$.

The EM algorithm iteratively alternates between its E-step, in which the
expectations of the indicators $\chi_{ijm}$ and $\chi'_{ijkm}$ are computed
for fixed parameter estimates, and its M-step, in which these expectations
are used to maximize the expectation of the log-likelihood function
$\ell_{\rm hid}$ and to obtain new parameter estimates.  Details of the EM
algorithm, including the choice of starting solutions and the bootstrap
procedure for confidence intervals, are given in the Appendix.

\section{Results}
\label{sec:results}

\subsection{Tree reconstruction}

\begin{figure}
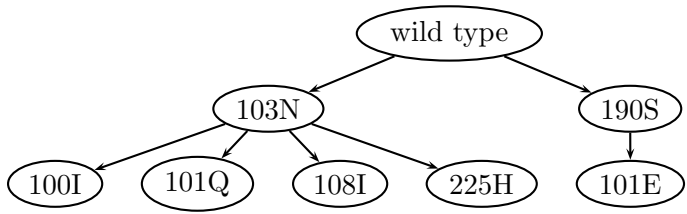

\centering
\tree{\root{wild type}}{
        \tree{\node{103N}}{
                \node{100I}
                \node{101Q}
                \node{108I}
                \node{225H}
                }
        \tree{\node{190S}}{
                \node{101E}
                }
        }
\caption{Mutagenetic tree model for the development of 
  resistance to efavirenz in the HIV-1 reverse transcriptase.
  Vertices are labeled with amino acid substitutions.}
\label{fig:mtree}
\end{figure}

Employing the cross-sectional tree reconstruction method yielded the
mutagenetic tree on 8 vertices displayed in Figure~\ref{fig:mtree} with
parameters $\lambda$ shown in Figure~\ref{fig:eps}(a). In this mutagenetic
tree model, predominantly mutation 103N initiates the development of
efavirenz resistance and is followed by 100I, 101Q, 108I, or 225H. A
parallel but strongly retarded pathway involves mutation 190S followed by
101E. Based on this cross-sectional analysis, the expected waiting time for
mutation 103N to occur is estimated to $\hat{\lambda}_{\rm 103N}^{-1} =
3.4$ weeks of efavirenz combination therapy. By contrast, mutation 190S is
estimated to occur after 113 weeks.

The mutagenetic tree model has 51 compatible states. The transition
matrix of the corresponding HMM has 588 nonzero entries out of $51^2 =
2601$ (cf.\ Appendix).

\subsection{Inferred population states}

Using the fixed tree topology, we applied the EM algorithm to the
longitudinal clonal data. The Viterbi algorithm was used to obtain maximum
a posteriori estimates of the hidden variables, i.e.\ of the viral
population states, resulting in one Viterbi path for each patient.  Given
the model parameters, the Viterbi path is the most probable sequence of
states to explain both the progression in time of the virus population
along the lattice of compatible states and the observed clones at the
respective time points.  For example, the Viterbi path for patient~22 is illustrated in
Table~\ref{tab:P22}.  We estimated mutation 103N to enter the virus
population of this patient at week 48 and mutation 225H to be introduced
after 70 weeks of therapy. Mutation 103N had already been present in 2 out
of 16 clones before onset of therapy (week 0), but was most likely not
established in the population at this time, because viral replication was
suppressed for 48 more weeks \citep[Fig.~1]{Bacheler2000}.  Likewise,
mutation 225H appears first at week 59 in 1 out of 7 clones, 11 weeks
before its estimated fixation.  Conversely, we estimated the introduction
of this mutation at week 70 despite the fact that 4 out of 8 cloned did not
yet harbor it at this time.

\subsection{Rates of progression}

\begin{figure*}[!tpb]
\centerline{
\begin{tabular}{ccc}
\includegraphics[width=5.5cm]{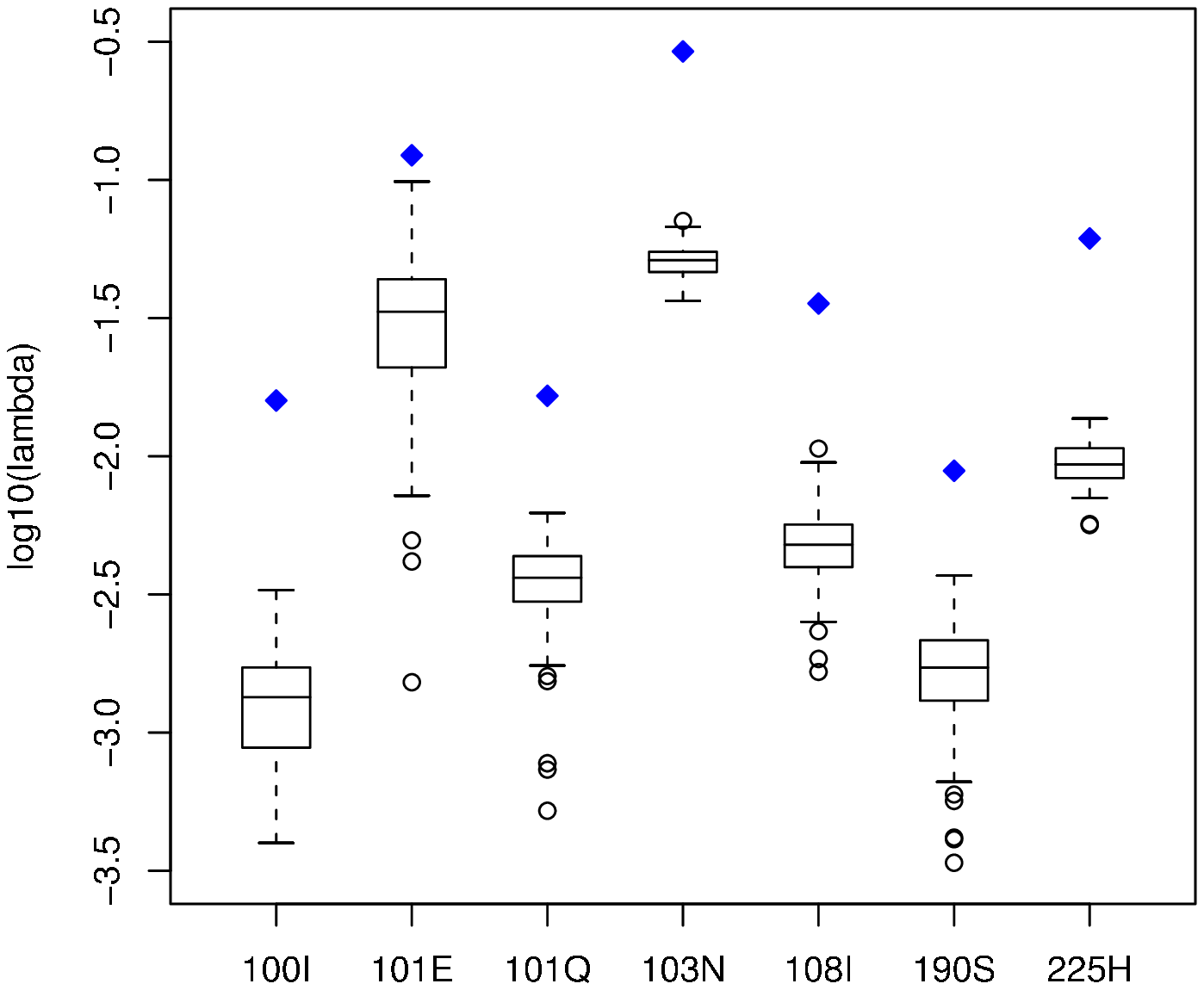} &
\includegraphics[width=5.5cm]{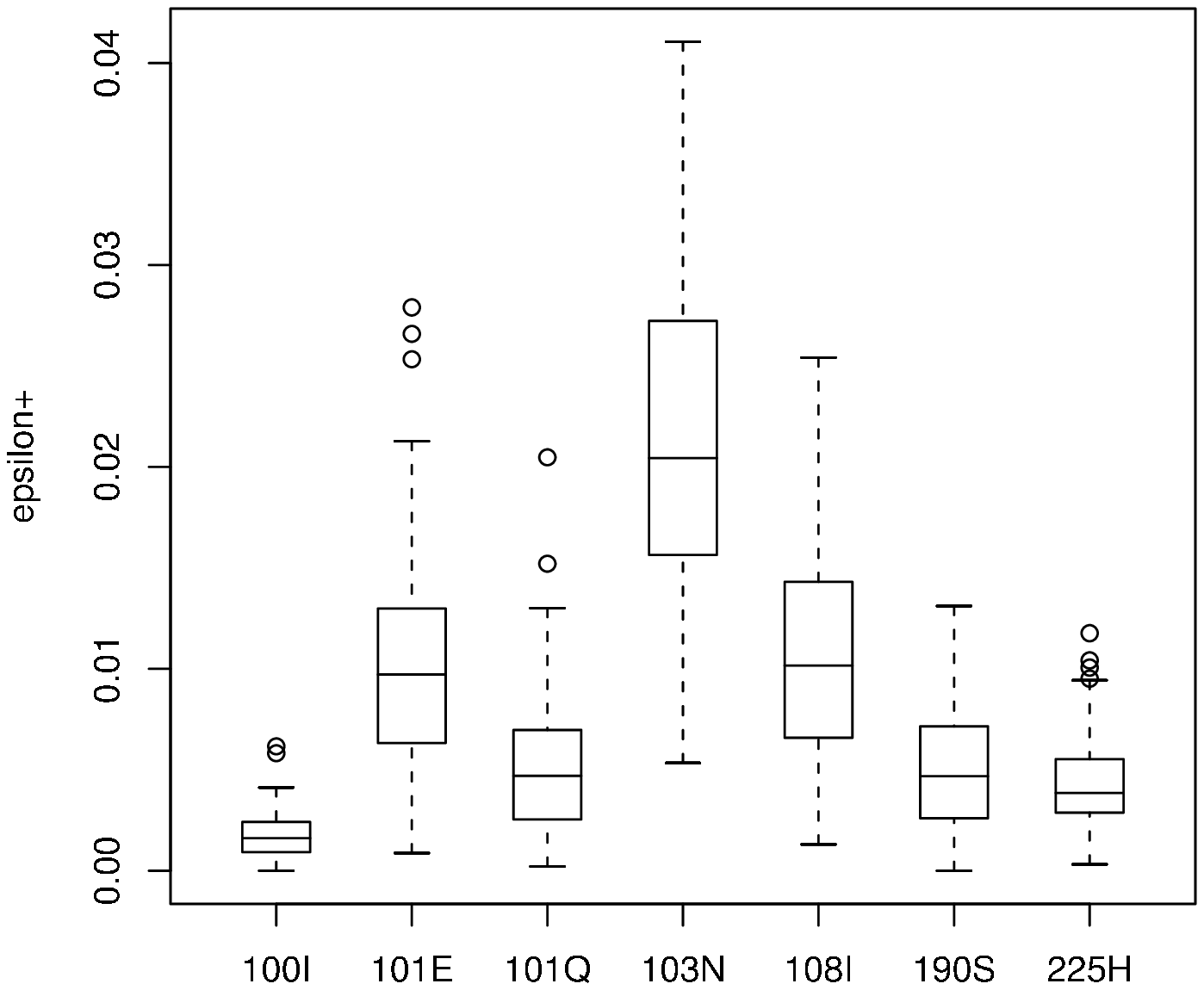} &
\includegraphics[width=5.5cm]{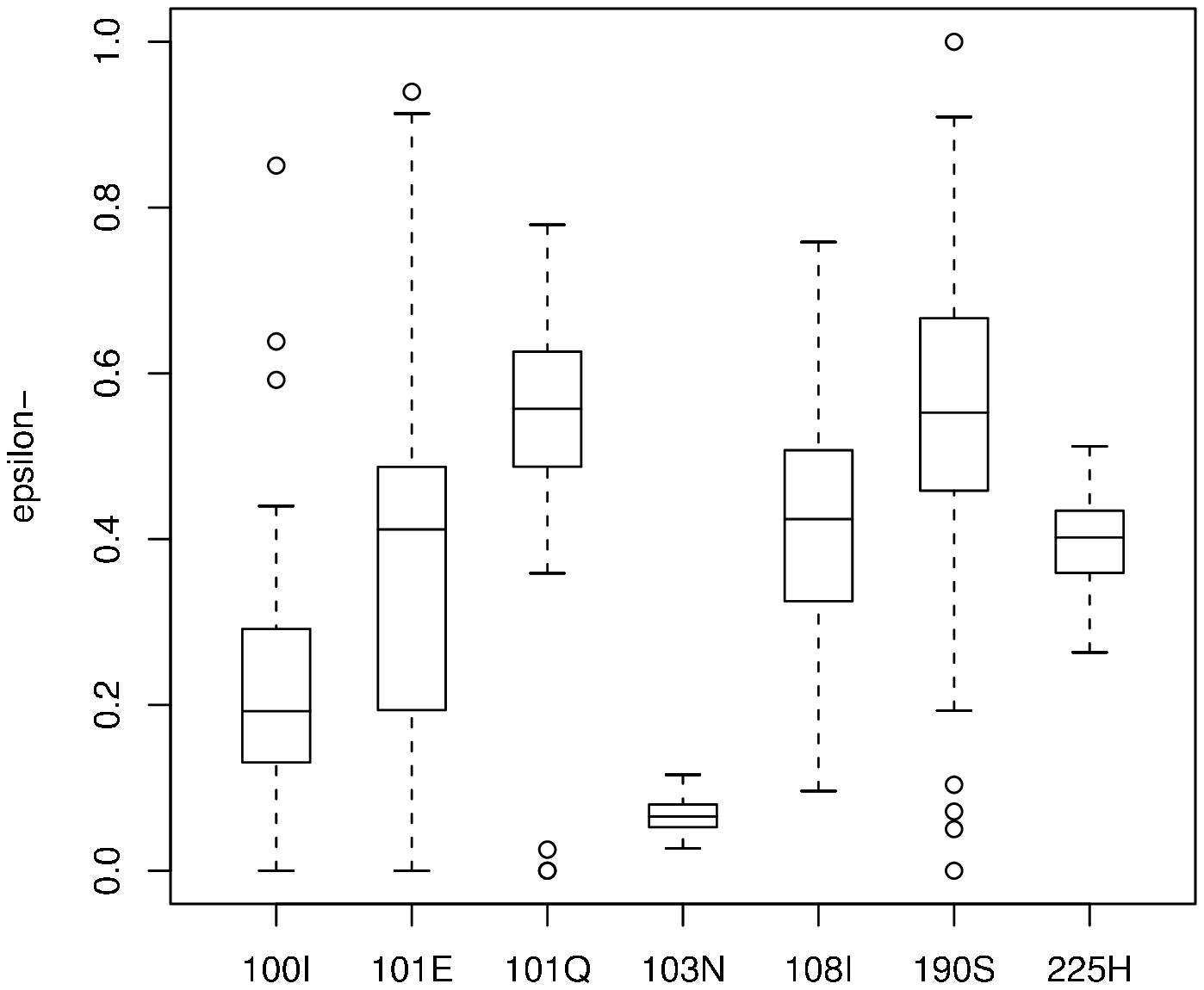} \\[-4ex]
(a) & (b) & (c)
\end{tabular}
}
\caption{Bootstrap estimates of the parameters of the mtree-HMM for the
  development of efavirenz resistance based on the tree in
  Figure~\ref{fig:mtree}. (a) Progression rates $\lambda$ of mutations,
  reported on a log scale. Diamonds indicate the cross-sectional
  estimate. (b) False positive rates $\varepsilon^+$. (c) False negative
  rates $\varepsilon^-$.} 
\label{fig:eps}
\end{figure*}

The parameters $\lambda$ are conditional rates of fixation of mutations
in the population, associated with the edges of the mutagenetic tree.  Estimation
of these rates from cross-sectional data depends on the assumption of
independent observations and of an exponentially distributed sampling time
(Equation~\ref{eqn:lambda} in the Appendix), both of which are not met with
the present data set ($p<10^{-15}$, Kolmogorov-Smirnov test).  Estimates of
$\lambda$ in the mtree-HMM are shown in Figure~\ref{fig:eps}(a).

Mutation 103N is by far the earliest mutation to occur with an expected
waiting time of 19 weeks, as compared to 190S with 478 weeks.  Subsequent
to 103N, mutation 225H is most likely to occur next, followed by 108I,
101Q, and 100I. Mutation 101E appears shortly after 190S. These findings
are in accordance with {\it in vitro} measurements of efavirenz resistance
\citep{Bacheler2001}. Engineered single amino acid substitutions result in
36-fold and 97-fold reduced efavirenz susceptibility for 103N and 190S,
respectively, as compared to the wild type. By contrast, all other
mutations cause significantly less efavirenz resistance on their own (1.2-
to 24-fold, median 5.6).  Remarkably however, all of the double mutants
103N+100I, 103N+101Q, 103N+108I, 103N+225H as well as the triple mutant
103N+108I+225H were measured highly resistant to efavirenz (84- to
2,400-fold). Thus, phenotypic resistance increases along the mutagenetic
tree, and the ordered accumulation appears to result from a specific
phenotypic gradient.

Comparing the cross-sectional estimates of $\lambda$ to their longitudinal
counterparts obtained from bootstrapping reveals consistent overestimation
of progression rates by the cross-sectional approach
(Figure~\ref{fig:eps}(a)).  This effect may be due to the high number of
clones that have been sequenced prior to therapy start at week 0, which
leads to an estimated sampling time of only $22$ weeks in the
cross-sectional approach.

\subsection{Clonal variation}

The variation of clones is modeled by mutation-specific parameters
$\varepsilon^+_m$ and $\varepsilon^-_m$ that denote the probability of a
false positive and a false negative mutation, respectively.
Figures~\ref{fig:eps}(b) and \ref{fig:eps}(c) show the bootstrap estimates
of these parameters. False positives occur when a mutation is not yet fixed
into the population, but is already present on single clones. We found very
low false positive rates ($\mbox{mean} < 0.02$) for all mutations.

We obtained higher estimates for $\varepsilon^-$, the rates of false
negatives. False negatives occur when a mutation is estimated to be fixed
into the virus population, but does not appear in subsequent clones. This
happens more frequently, because the model assumes non-reversibility of
mutations in the population.  The false negative rate of 103N was by far
the lowest, presumably due to its strong selective advantage both as a
single mutation and in combination with others.  Mutations 190S and 101E
showed high false negative rates but these estimates were also subject to a
considerable variance.  
In the case of 190S and also 101Q, the
high false negative rate is not as surprising given that the randomization
tests in Section~\ref{sec:statistical-tests} indicated that the assumption of
non-reversibility of substitution is less warranted for these two
mutations.

\section{Limitations and conclusions}
\label{sec:limit}

RNA viruses provide prominent examples of measurably evolving populations,
and precise predictions of the outcome of this evolutionary process are a
prerequisite for the rational design of antiretroviral treatment protocols.
Insight in the evolutionary process can be obtained from longitudinal
genomic data.  For analysis of such data, we have presented a mutagenetic
tree hidden Markov model.  We applied this model in an analysis of longitudinal clonal
sequence data on the evolution of drug resistance in HIV.  Our focus was on
estimating the order and rate of occurrence of seven single amino acid
changes that are associated with resistance to the drug efavirenz.

Our model of clonal variation assumes that the clones are possibly
erroneous copies of a fixed but unobserved population state.  Thus, we
regard the virus population as consisting of a single dominating strain and
a cloud of closely related mutants.  While this assumption is justified by
the strong selective pressure exerted by antiretroviral therapy, it does
not allow for explicit modeling of different lineages following different
evolutionary pathways. Such parallel developments were very rare in the
present data set. For example, out of the 7 viral clones derived from
patient 126 at week 59, 3 clones displayed the mutational pattern
103N-225H, and 5 mutations 190S-101E. The estimated population state was
the union of these four mutations, but none of the observed clones actually
displayed this pattern. In fact, these parallel lineages contribute to the
elevated false negative rates of mutations 101E and 190S.  Possible ways of
extending the mtree-HMM to account for this effect include modeling the
population state as more than one strain, and grouping clones into
different pathways.

Evolution in our model is assumed to result from mutation and selection. 
Incorporating recombination would alter our set-up
considerably. However, the mutations we considered lie no more than 125
codons, or 375 nucleotides, apart. Given the HIV genome of length ~10,000
base pairs, recombination is unlikely to occur in this small region, but
cannot be ruled out entirely.

Finally, we emphasize that throughout we have ignored the interval
censoring inherent in the data and  assumed that mutations occur at
the time they are observed. However, ignoring the interval censoring is
unlikely to have a strong effect on the parameter estimates because of the
small interval lengths. In a related analysis of data with the same time
interval structure, \citet{Foulkes2003b} have found no qualitative and
little quantitative difference when using the midpoint of time intervals as
the time at which mutations occur. Nevertheless, modeling interval
censoring will become more important for larger time intervals, for example
due to longer periods of suppressed virus replication below detectable
limits.

\section*{Acknowledgments}

Niko Beerenwinkel is supported by Deutsche Forschungsgemeinschaft (DFG)
under grant No.\ BE~3217/1-1.  Mathias Drton acknowledges support from NIH
grant R01-HG02362-03 and NSF grant DMS-0505612.

\bibliographystyle{abbrvnat}


\begin{appendix}
\section{Expectation maximization algorithm}

We provide here a detailed description of the EM algorithm from Section~\ref{sec:statinf}
that is used to estimate the parameters of the mtree-HMM.

\subsection{M-step}

The log-likelihood function $\ell_{\rm hid}$ of the hidden model
decomposes into a sum
involving only $\lambda$ and another sum involving only
$\varepsilon^+$ and $\varepsilon^-$. Let
\[
u_{ijm,a} = \Prob(x_{i,j-1,m}=0,\; 
x_{ij\pa(m)}=1,\; x_{ijm}=a \mid Y),
\]
be the conditional expectation of $\chi_{ijm}(a)$ given the clonal
observations $Y$.  For maximum likelihood estimation of $\lambda$, we
have to maximize the function
\begin{equation*}
\sum_{i\in [N]} \sum_{m\in [M]} \sum_{j=1}^{J_i} \big[ - u_{ijm,0} \,
\lambda_m \, (t_{ij} - t_{i,j-1})  + u_{ijm,1} \, \log\big(1 -
e^{\lambda_m (t_{ij} - t_{i,j-1})}\big) \big].
\end{equation*}
This task is complicated by the $\log$ terms, but can be 
accomplished numerically by simple general optimization methods, 
such as the simplex method of \citet{Nelder1965}.

For estimating $\varepsilon^+$ and $\varepsilon^-$ let
\[
u'_{m,ab} = \sum_{i\in [N]} \sum_{j=1}^{J_i} \sum_{k\in
  [K_{ij}]} \Pr(x_{ijm} = a, \; y_{ijkm} = b \mid Y)
\]
be the conditional expectation of the corresponding sum of
$\chi'_{ijkm}(a,b)$ given the clonal observations $Y$.  Then the maximum
likelihood estimates are
\[
   \hat\varepsilon^+_m = \frac{u'_{m,01}}{u'_{m,00} + u'_{m,01}},
   \quad
   \hat\varepsilon^-_m = \frac{u'_{m,10}}{u'_{m,10} + u'_{m,11}},
   \quad
   m \in [M].
\]

\subsection{E-step}

The mtree-HMM is a submodel of a standard HMM with hidden
state space equal to the set $\cC(T)$ of states compatible with the
mutagenetic tree $T$.  Therefore, we can compute the conditional
expectations $u_{ijv,a}$ and $u'_{m,ab}$ by applying, separately for
each patient, the forward-backward equations of the standard HMM
\citep{Durbin1998,Hallgrimsdottir2005}.  In the standard HMM only certain
transitions between compatible states can occur with non-zero
probability, namely those that respect the partial order ''$\succeq$''
of the lattice induced by $T$.
The probability $\Prob(X_{ij}=\bar x\mid X_{i,j-1}=x)$
 of transitioning from state $x\in\cC(T)$
to state $\bar x\in\cC(T)$ is given by
\begin{equation*}
\prod_{m\in [M]}
\theta_{ij}(\lambda_m)_{(x_{m},\bar x_{\pa(m)}),\bar x_{m}}
= \begin{cases}
  \frac{\Prob(W=\bar x)}{\zeta}  
  & \mbox{if $\bar x \succeq x$}, \\
  0  & \mbox{else,}
\end{cases}
\end{equation*}
where $\zeta = \sum_{x' \succeq x} \Prob(W=x')$ and
$W\in \cI$ is a random vector distributed according to the timed
mutagenetic tree model based on the tree $T$ and the time interval
$\Delta t=t_{ij}-t_{i,j-1}$.  The number of compatible states
depends on the topology of $T$ but is
bounded by $M+1\le |\cC(T)|\le 2^M$.  Thus the number of non-zero
transition probabilities also depends on the topology of $T$.
For example, the transition matrix arising from the tree $T$ 
shown in Figure \ref{fig:lattice}(a) is of the form
\[
\bordermatrix{
     & 0000 & 1000 & 1010 & 1100 & 1011 & 1110 & 1111 \cr
0000 &   *  &   *  &   *  &   *  &   *  &   *  &   *  \cr
1000 &   0  &   *  &   *  &   *  &   *  &   *  &   *  \cr
1010 &   0  &   0  &   *  &   0  &   *  &   *  &   *  \cr
1100 &   0  &   0  &   0  &   *  &   0  &   *  &   *  \cr
1011 &   0  &   0  &   0  &   0  &   *  &   0  &   *  \cr
1110 &   0  &   0  &   0  &   0  &   0  &   *  &   *  \cr  
1111 &   0  &   0  &   0  &   0  &   0  &   0  &   *  \cr  
},
\]
where asterisks indicate entries that are not restricted to zero.  

The remaining structure of the standard HMM is
determined by the probability of emitting clones
$y_{ij1},\dots,y_{ijK_{ij}}$ from a hidden state $x \in \cC(T)$, which
is equal to
\[
\prod_{m\in [M]} 
\prod_{k\in [K_{ij}]} 
\theta'(\varepsilon^+_m,\varepsilon^-_m)_{x_{m},y_{ijkm}}.
\]

Applying the forward-backward equations to this HMM yields,
for each patient $i$ and time point $t_{ij}$, the conditional
expectation of the number of transitions $U_{ij,x\bar x}$ from state
$x\in\cC(T)$ to state $\bar x\in\cC(T)$, and the conditional
expectation of the number of emissions $U'_{ij,x y}$ of clone
$y\in\cI$ from state $x\in\cC(T)$.  From these conditional
expectations, we can obtain  the quantities
\[
\begin{split}
u_{ijm,a} &= 
\sum_{ \{ (x,\bar x) \in \cC(T) \times \cC(T) \;:\; \atop
  x_m = 0, \; \bar x_{\pa(m)} = 1, \; \bar x_m = a \} } U_{ij,x\bar
  x},\\
u'_{m,ab} &= 
\sum_{i\in [N]}\sum_{m\in [M]} \sum_{j=1}^{J_i}
\sum_{ \{ (x,y) \in \cC(T) \times \cI \;:\; \atop
  x_m = a, \; y_m = b \} } U'_{ij,xy}
\end{split}
\]
that are needed in the M-step of the EM algorithm for the mtree-HMM.

If several vertices of the mutagenetic tree share the root vertex 0 
as parent, additional computational saving is possible by employing 
that variables in different subtrees rooted at 0 are mutually independent.
In this case several standard HMMs with smaller state spaces can
replace the one standard HMM.

\subsection{Starting solutions}

As initial parameter values in the mtree-HMM we used $0.1$ for all
error probabilities $\varepsilon^+_m$ and $\varepsilon^-_m$. The
initial rates $\lambda$ were derived from the cross-sectional approach
by assuming an exponential distribution with rate $\lambda_T$ for the
sampling times. Under this assumption $\lambda$ can be computed from
\begin{equation} \label{eqn:lambda}
   \vartheta^m_{11} = \frac{\lambda_m}{\lambda_m + \lambda_T},
      \quad m \in [M],
\end{equation}
where both $\vartheta^m_{11}$ and $\lambda_T$ were estimated directly
from the cross-sectional data \citep{Rahnenfuehrer2005}.

\subsection{Confidence intervals}

We obtained confidence intervals for all model parameters $\lambda$,
$\varepsilon^+$, and $\varepsilon^-$ from 100 bootstrap runs of the EM
algorithm. Resampling of the data was performed in two steps.  First,
a set of 163 patients was drawn at random. Second, for each patient
and each time point, the corresponding set of clones was used for
resampling the respective number of clones. Thus, the number of
patients, the structure of sampling times, and the number of clones
equaled those of the original data set in all bootstrap runs.

\vspace{1cm}

\end{appendix}

\end{document}